\documentclass[12pt]{article}
\usepackage[dvips]{epsfig}
\begin{document}
\begin{flushright}
OHSTPY-HEP-T-98-014 \\
hep-th/9808120
\end{flushright}
\vspace{20mm}
\begin{center}
{\LARGE DLCQ Bound States of ${\cal N} =(2,2)$  Super-Yang-Mills 
at Finite and Large $N$ }
\\
\vspace{20mm}
{\bf F.Antonuccio${}^{(1)}$, H.C.Pauli${}^{(2)}$,
  S.Pinsky${}^{(1)}$ and 
S.Tsujimaru${}^{(2)}$} \\
\vspace{4mm}
{\em ${}^{(1)}$Department of Physics,\\ The Ohio State University,\\ Columbus,
OH 43210, USA\\
\vspace{4mm} and \\
\vspace{4mm}
${}^{(2)}$Max-Planck-Institut f\"{u}r Kernphysik, \\ 69029 Heidelberg, Germany}

\end{center}
\vspace{10mm}

\begin{abstract}

We consider the $1+1$ dimensional ${\cal N} = (2,2)$
supersymmetric matrix model which is 
obtained by dimensionally reducing 
${\cal N} = 1$  super Yang-Mills from four to
two dimensions.
The gauge groups we consider are  
U($N$) and SU($N$), where $N$ is finite but arbitrary. 
We adopt light-cone coordinates, and choose to work in the light-cone gauge.
Quantizing this theory via Discretized Light-Cone Quantization (DLCQ) 
introduces an integer, $K$, which restricts
the light-cone momentum-fraction of constituent quanta to be
 integer multiples of
$1/K$. Solutions to the DLCQ bound state equations are 
obtained for $2 \leq K \leq 6$ by 
discretizing the light-cone supercharges,
which results in a supersymmetric spectrum.
Our numerical results imply the existence of normalizable
massless states in the continuum
limit $K \rightarrow \infty$, and 
therefore the absence of a mass gap.
The low energy spectrum is dominated
by string-like (or many parton) states.
Our results are consistent
with the claim that the theory is in a screening phase. 

\end{abstract}
\newpage

\baselineskip .25in

\section{Introduction}
Supersymmetric gauge theories in low dimensions 
have been shown to be related to non-perturbative objects
in M/string theory \cite{witt95}, and are therefore of
particular interest nowadays. More dramatically, there is a
growing body of evidence suggesting that 
gauged matrix models in $0+1$ and $1+1$ dimensions
may offer a non-perturbative formulation of string theory 
\cite{bfss97,dvv}. There is also a suggestion that large $N$ gauge
theories in various dimensions may be related to theories with gravity
\cite{mald97}.

It is therefore interesting to study directly the non-perturbative
properties of a class of supersymmetric matrix models at finite
and large $N$, where $N$ is the number of gauge colors.
In previous work \cite{alp98,alpII98}, we focused on two dimensional
matrix models, since the numerical technique of Discrete Light-Cone
Quantization (hereafter `DLCQ' \cite{pb85}) may be 
implemented to determine
bound state wave functions and masses.

In this work, we extend these investigations by 
solving the DLCQ bound state equations for
a two dimensional
supersymmetric matrix model with ${\cal N} = (2,2)$
supersymmetry. Such a theory may be obtained by
dimensionally reducing ${\cal N}=1$ super 
Yang-Mills from four to two space-time dimensions 
\cite{bss77}. Various studies related to this model can
be found in the literature \cite{witt93,dorey}, 
and it has recently been shown
that this theory exhibits a screening phase \cite{armoni}.
 
The content of this paper is divided up as follows.
In Section \ref{dlcqformulation}, we formulate
the ${\cal N}=(2,2)$ supersymmetric matrix model 
in the light-cone frame, and quantize the theory by
imposing canonical (anti)commutation relations for
fermions and bosons respectively. In Section
\ref{numericalresults}, we briefly describe 
the DLCQ numerical procedure, and present 
our numerical results for the bound state spectrum.
The structure of bound state wave functions is also 
discussed.
A summary of our work appears in the discussion of
Section \ref{conclusions}. Details of the underlying
four dimensional ${\cal N}=1$ super Yang-Mills theory
(i.e. before dimensional reduction) can be found
in Appendix \ref{ymills4}.

\section{Light-Cone Quantization and DLCQ at Finite $N$}
\label{dlcqformulation}
The two dimensional ${\cal N} = (2,2)$
supersymmetric gauge theory we are interested in
may be formally obtained 
by dimensionally reducing ${\cal N}=1$
super Yang-Mills from four to two dimensions. For the sake of completeness,
we review the underlying four dimensional Yang-Mills theory
in Appendix \ref{ymills4}.

Dimensional reduction of the four dimensional Yang-Mills
action (\ref{LCversion}) given in  Appendix \ref{ymills4}
is carried out by
stipulating that all fields are independent of the {\em two} 
transverse
coordinates\footnote{The space-time points
in four dimensional Minkowski space are parameterized, as usual,
by coordinates
$(x^0,x^1,x^2,x^3)$.} 
$x^I=-x_I$, $I=1,2$. 
We may therefore assume
that the fields depend only on the light-cone variables $\sigma^{\pm}
= \frac{1}{\sqrt{2}}(x^0 \pm x^3)$. 
The resulting two dimensional theory may be described 
by the action
\begin{eqnarray}
S_{1+1}^{LC} & = & \int d\sigma^+ d\sigma^- \hspace{1mm}
 \mbox{tr} \left( \frac{1}{2}D_\alpha X_I D^\alpha X_I + \frac{g^2}{4}
            [X_I,X_J]^2 - \frac{1}{4} F_{\alpha \beta} F^{\alpha \beta} 
 \right. \nonumber \\
& & \hspace{20mm}
+ \hspace{1mm}
{\rm i} \theta_R^T D_+ \theta_R +   {\rm i}\theta_L^T D_- \theta_L 
    + \sqrt{2}g\theta_L^T \epsilon_2 \beta_I
[X_I,\theta_R] \left. \frac{}{} \right),  
\label{LCversionreduced}
\end{eqnarray}
where the repeated indices $\alpha,\beta$ are summed over light-cone
indices $\pm$, and $I,J$ are summed over transverse
indices $1,2$.
The two scalar fields $X_I(\sigma^+,\sigma^-)$ represent
$N \times N$ Hermitian matrix-valued fields, and are 
remnants of
the transverse components of the
four dimensional gauge field $A_\mu$, while $A_{\pm}(\sigma^+,\sigma^-)$ 
are the
light-cone gauge field components of the residual 
two dimensional U($N$) or SU($N$) gauge symmetry. 
The two component spinors $\theta_R$ and
$\theta_L$ are remnants of
the right-moving and left-moving
projections of a four component real spinor in the four dimensional
theory. The components of  $\theta_R$ and $\theta_L$
transform in the adjoint representation of the gauge group.  
  $F_{\alpha \beta} =  
\partial_{\alpha} A_\beta - \partial_\beta A_\alpha
    +{\rm i}g[A_\alpha, A_\beta]$ is the two dimensional
gauge field curvature tensor, while
$D_\alpha =  \partial_\alpha + {\rm i}g[A_\alpha,\cdot]$ is the covariant
derivative for the (adjoint) spinor fields.
The two $2 \times 2$ real symmetric matrices
$\beta_I$, and anti-symmetric matrix $\epsilon_2$,
are defined in Appendix \ref{ymills4}.

Since we are working in the light-cone frame, it is natural
to adopt the light-cone gauge $A_- = 0$. With
this gauge choice, the action (\ref{LCversionreduced}) becomes   
\begin{eqnarray}
{\tilde S}_{1+1}^{LC}&=&
\int d\sigma^+d\sigma^- {\rm {tr}} \Bigg(\partial_+X_I\partial_-X_I +
{\rm i} 
\theta_R^T\partial_+ \theta_R + {\rm i}\theta_L^T\partial_- \theta_L 
\nonumber\\
&+&\frac{1}{2}(\partial_-A_+)^2 +gA_+J^+ 
+\sqrt{2}g \theta_L^T \epsilon_2 \beta_I [X_I, \theta_R ] 
+\frac{g^2}{4}[X_I, X_J ]^2\Bigg), 
\label{EQ6}
\end{eqnarray}
where $J^+ ={\rm i}[X_I, \partial_-X_I]+2\theta_R^T\theta_R$ 
is the longitudinal
momentum current.
The (Euler-Lagrange) equations of motion for the $A_+$
and $\theta_L$ fields are now
 \begin{eqnarray}
&&\partial_-^2A_+=gJ^+, \label{firstc}\\
&& \sqrt2 {\rm i}\partial_-\theta_L= -g\epsilon_2 \beta_I [X_I,\theta_R].
\label{secondc}  
\end{eqnarray}
These are evidently constraint equations, since
they are independent of the light-cone time $\sigma^+$.
The ``zero mode'' of the constraints above provide
us with the conditions
\begin{equation}
\int d\sigma^- J^+=0, \mbox{     and      } 
\int d\sigma^- \epsilon_2 \beta_I [X_I,\theta_R] =0,
\label{EQ4}
\end{equation}    
which will be imposed on the Fock space 
to select the physical states in the quantum theory. 
The first constraint above is well known in the literature,
and projects out the colorless states in
the quantized theory\cite{dak93}. The second (fermionic) constraint is
perhaps lesser well known, but certainly provides  non-trivial
relations governing the small-$x$ behavior of light-cone 
wave functions\footnote{If we introduce a mass term, such
relations become crucial in establishing finiteness 
conditions. See \cite{abd97}, for example.} \cite{abd97}. 
  
At any rate, equations (\ref{firstc}),(\ref{secondc}) permit one
to eliminate the non-dynamical fields $A_+$ and $\theta_L$  
in the expression for the light-cone Hamiltonian $P^-$, 
which is a particular feature of
light-cone gauge theories. There are no ghosts.
We may therefore write down explicit  expressions for the
light-cone momentum $P^+$ and Hamiltonian $P^-$ exclusively
in terms of the physical degrees of freedom represented 
by the two scalar fields 
$X_I$ and two-component spinor $\theta_R$:
\begin{eqnarray}
P^+&=&\int d\sigma^- \hspace{1mm}
\mbox{tr}
\left( \partial_-X_I\partial_-X_I+{\rm i}
\theta_R^T \partial_-\theta_R \right), \label{P+}
\\  
P^- &=&g^2 
\int d\sigma^- {\rm {tr}}\Bigg(-\frac{1}{2} J^+\frac{1}{\partial_-^2}J^+
-\frac{1}{4}[X_I, X_J ]^2 \nonumber \\ 
&&\hspace{15mm}+\frac{{\rm i}}{2} 
(\epsilon_2 \beta_I [X_I, \theta_R])^T
\frac{1}{\partial_-} \epsilon_2 \beta_J [X_J, \theta_R]\Bigg). 
\label{P-}
\end{eqnarray}
The light-cone Hamiltonian propagates a given field configuration
in light-cone time $\sigma^+$, and contains all the non-trivial
dynamics of the interacting field theory. 

Let us denote the two components of the spinor $\theta_R$ by
the fermion fields $u^{\alpha}$, $\alpha=1,2$. 
Then,
in terms of their Fourier modes, the physical fields 
may be expanded at light-cone time $\sigma^+=0$ to give\footnote{
The symbol $\dagger$ denotes quantum conjugation, and does not
transpose matrix indices.} 
\begin{eqnarray}
&&X^I_{pq}(\sigma^-)= \frac{1}{\sqrt{2\pi}}
\int_{0}^{\infty}\frac{dk^+}{\sqrt{2 k^+}}\Big(a^I_{pq}(k^+)
e^{-{\rm i}k^+\sigma^-}
+ {a^I_{qp}}^{\dagger}(k^+)e^{{\rm i}k^+\sigma^-}\Big), 
\hspace{4mm} I=1,2; \hspace{3mm} \label{Xexp}\\
&&u^{\alpha}_{pq}(\sigma^-)=\frac{1}{\sqrt{2 \pi}}\int_0^{\infty}
\frac{dk^+}{\sqrt{2}} 
\Big(b^{\alpha}_{pq}(k^+)e^{-{\rm i}k^+\sigma^-}
+ {b^\alpha_{qp}}^{\dagger}(k^+)e^{{\rm i}k^+\sigma^-}\Big), 
\hspace{4mm} \alpha=1,2. \label{uexp}
\end{eqnarray}
For the gauge group U($N$), the (anti)commutation relations take the form
\begin{eqnarray}
&&[a^I_{pq}(k^+), {a^J_{rs}}^{\dagger}(k'^+)]=
\delta^{IJ}\delta_{pr}\delta_{qs} 
\delta(k^+- k'^+), \\
&&\{ b^{\alpha}_{pq}(k^+), {b^{\beta}_{rs}}^{\dagger}(k'^+)\}=
\delta^{\alpha\beta}
\delta_{pr}\delta_{qs}\delta(k^+- k'^+),
\end{eqnarray}
while for SU($N$), we have the corresponding relations 
\begin{eqnarray}
&&[a^I_{pq}(k^+), {a^J_{rs}}^{\dagger}(k'^+)]=
\delta^{IJ}(\delta_{pr}\delta_{qs} - \frac{1}{N}\delta_{pq} \delta_{rs}) 
\delta(k^+- k'^+), \\
&&\{ b^{\alpha}_{pq}(k^+), {b^{\beta}_{rs}}^{\dagger}(k'^+)\}=
\delta^{\alpha\beta}
(\delta_{pr}\delta_{qs}
- \frac{1}{N}\delta_{pq} \delta_{rs})
\delta(k^+- k'^+).
\end{eqnarray}
An important simplification of the light-cone quantization is that 
the light-cone vacuum  is the Fock vacuum $\vert 0 \rangle$, defined by   
\begin{equation}
a^I_{pq}(k^+)\vert 0 \rangle =b^{\alpha}_{pq}(k^+)\vert 0 \rangle=0, 
\end{equation}
for all positive longitudinal momenta $k^+ > 0$. 
We therefore have $P^+\vert 0 \rangle= P^-\vert 0 \rangle=0$.
  
The ``charge-neutrality'' condition (first integral constraint
from (\ref{EQ4})) requires that all the color indices must be 
contracted for physical states.
Thus physical states are formed by color traces of the 
boson and fermion creation operators 
${a^I}^{\dagger},{b^{\alpha}}^{\dagger}$
acting on the light-cone vacuum.  A single trace of these
creation operators may be identified
as a single closed string, where each creation operator
(or `parton'), carrying some longitudinal
momentum $k^+$,  represents a
 `bit' of the string. A product of traced operators
is then a multiple string state, and the quantity $1/N$
is analogous to a string coupling constant \cite{thorn}.

\medskip

At this point, we may determine explicit expressions
for the quantized light-cone operators $P^{\pm}$ by substituting
the mode expansions (\ref{Xexp}),(\ref{uexp}) into 
equations (\ref{P+}),(\ref{P-}). The mass operator
$M^2 \equiv 2 P^+ P^-$ may then be diagonalized to solve for the bound 
state mass spectrum.
However, as was pointed out in \cite{sakai95},
it is more convenient to determine the quantized expressions
for the supercharges, since this leads to a regularization prescription
for $P^-$ that preserves supersymmetry even in the discretized theory.

In order to elaborate upon this last remark, first note that
the continuum theory possesses four supercharges,
which may be derived from the dimensionally reduced 
form of the four dimensional ${\cal N} = 1$ supercurrent
(eqn (\ref{supercurrent}) in Appendix \ref{ymills4}):
\begin{eqnarray}
 Q^+_{\alpha} & = & 2^{5/4} \int_{-\infty}^{\infty} d\sigma^- \hspace{1mm}
\mbox{tr} \left( \partial_- X_I \cdot \beta_{I \alpha \eta} \cdot u_{\eta} 
     \right) \label{Q+}\\
 Q^-_{\alpha} & = & g \int_{-\infty}^{\infty} d\sigma^- \hspace{1mm}
\mbox{tr} \left( -2^{3/4} \cdot J^+  \frac{1}{\partial_-} 
              {\epsilon_2}_{\alpha \eta} u_{\eta} +  
      2^{-1/4} {\rm i} [X_I,X_J] \cdot 
(\beta_I \beta_J \epsilon_2)_{\alpha \eta} \cdot u_{\eta}  \right),
\label{Q-}
\end{eqnarray}
where $\alpha=1,2$, and repeated indices are 
summed. The two $2 \times 2$ real symmetric
matrices $\beta_I$ are discussed in Appendix \ref{ymills4}. By explicit
calculation or otherwise, these charges satisfy the following
relations\footnote{Surface terms contributing to the central charge 
are assumed to be vanishing.}:
\begin{eqnarray}
       \{ Q^+_{\alpha}, Q^+_{\beta} \} & = & 
   \delta_{\alpha \beta} \cdot 2\sqrt{2} P^+ 
\label{superQplus} \\
 \{ Q^-_{\alpha}, Q^-_{\beta} \} & = & 
   \delta_{\alpha \beta} \cdot 2\sqrt{2} P^-  \label{superQminus} \\
\{ Q^-_{\alpha}, Q^+_{\beta} \} & = & 0, \hspace{5mm} \alpha,\beta=1,2. 
\label{crossterm} 
\end{eqnarray}
If we substitute the mode expansions (\ref{Xexp}),(\ref{uexp}) 
into equations (\ref{Q+}),(\ref{Q-}) 
for the light-cone supercharges $Q^{\pm}_{\alpha}$,
we obtain the following `momentum representations' for these charges:
\begin{eqnarray}
Q^+_{\alpha} & = & 2^{1/4} {\rm i} \int_0^{\infty}
 dk \hspace{1mm} \sqrt{k} \cdot \beta_{I \alpha \eta} \cdot 
\left( a^{\dagger}_{Iij}(k) b_{\eta ij}(k) -
       b^{\dagger}_{\eta ij}(k) a_{I ij}(k) \right),
\label{Qplus}
\end{eqnarray}
and 
\begin{eqnarray}
\lefteqn{ Q^-_{\alpha} = \frac{{\rm i} 2^{-1/4} g}{\sqrt{\pi}}
  \int_0^{\infty} dk_1 dk_2 dk_3 \hspace{1mm} 
\delta( k_1 + k_2 - k_3) \cdot (\epsilon_2)_{\alpha \eta} \cdot
 \left\{ \frac{}{} \right. }   & &  
\nonumber \\
& & 
\frac{1}{2\sqrt{k_1 k_2}} \left( \frac{k_2 - k_1}{k_3} \right)
\left[ b^{\dagger}_{\eta i j}(k_3)a_{I i m}(k_1)a_{I m j}(k_2) -
      a_{I i m}^{\dagger}(k_1)a_{I m j}^{\dagger}(k_2)
b_{\eta i j}(k_3) \right] \nonumber \\
& + & 
\frac{1}{2\sqrt{k_1 k_3}} \left( \frac{k_1 + k_3}{k_2} \right)
\left[ a^{\dagger}_{I i m}(k_1)b_{\eta m j}^{\dagger}(k_2)a_{I i j}(k_3) -
      a_{I i j}^{\dagger}(k_3)a_{I i m}(k_1)
b_{\eta m j}(k_2) \right] \nonumber \\
& + & 
\frac{1}{2\sqrt{k_2 k_3}} \left( \frac{k_2 + k_3}{k_1} \right)
\left[ a^{\dagger}_{I i j}(k_3)b_{\eta i m}(k_1)a_{I m j}(k_2) -
      b_{\eta i m}^{\dagger}(k_1)a_{I m j}^{\dagger}(k_2)
a_{I i j}(k_3) \right] \nonumber \\
& - & \frac{1}{k_1} \left[ 
b^{\dagger}_{\beta i j}(k_3)b_{\eta i m}(k_1)b_{\beta m j}(k_2) +
      b_{\eta i m}^{\dagger}(k_1) b_{\beta m j}^{\dagger}(k_2)
b_{\beta i j}(k_3) \right] \nonumber \\
& - & 
\frac{1}{k_2} \left[ 
b^{\dagger}_{\beta i j}(k_3)b_{\beta i m}(k_1)b_{\eta m j}(k_2) +
      b_{\beta i m}^{\dagger}(k_1) b_{\eta m j}^{\dagger}(k_2)
b_{\beta i j}(k_3) \right] \nonumber \\
& + &   
\frac{1}{k_3} \left[ 
b^{\dagger}_{\eta i j}(k_3)b_{\beta i m}(k_1)b_{\beta m j}(k_2) +
      b_{\beta i m}^{\dagger}(k_1) b_{\beta m j}^{\dagger}(k_2)
b_{\eta i j}(k_3) \right] \nonumber \\
& + & \hspace{8mm} 
[(\beta_I \beta_J - \beta_J \beta_I)\epsilon_2]_{\alpha \beta}
\times \left( \frac{}{} \right. \nonumber \\
& & \frac{1}{4\sqrt{k_1 k_2}}
\left[ b^{\dagger}_{\beta i j}(k_3)a_{I i m}(k_1)a_{J m j}(k_2) +
      a_{J i m}^{\dagger}(k_1)a_{I m j}^{\dagger}(k_2)
b_{\beta i j}(k_3) \right] \nonumber \\
& + & 
\frac{1}{4\sqrt{k_2 k_3}}
\left[ a^{\dagger}_{J i j}(k_3)b_{\beta i m}(k_1)a_{I m j}(k_2) +
      b_{\beta i m}^{\dagger}(k_1)a_{J m j}^{\dagger}(k_2)
a_{I i j}(k_3) \right] \nonumber \\
& + &
\frac{1}{4\sqrt{k_3 k_1}}
\left[ a^{\dagger}_{I i j}(k_3)a_{J i m}(k_1)b_{\beta m j}(k_2) +
      a_{I i m}^{\dagger}(k_1)b_{\beta m j}^{\dagger}(k_2)
a_{J i j}(k_3) \right] \left. \frac{}{} \right) \left. \frac{}{}
\right\}, \label{Qminus}
\end{eqnarray}
where repeated indices are always summed: $\alpha,\beta,\eta = 1,2$
(spinor indices), $I,J=1,2$ (SO(2) vector indices),
and $i,j,m=1,\dots , N$ (matrix indices).

In order to implement the DLCQ formulation\footnote{
It might be useful to consult \cite{sakai95,dak93,anp97,pin97}
for an elaboration of DLCQ in models
with adjoint fermions. An extensive list of references on DLCQ 
and light-cone field theories appears in the review \cite{bpp98}.}
of the bound state problem -- which is tantamount
to imposing periodic boundary conditions 
$\sigma^- \sim \sigma^- + 2 \pi R$ --
we simply restrict the momentum variable(s) 
appearing in the expressions for $Q^{\pm}_{\alpha}$ 
(equations (\ref{Qplus}),(\ref{Qminus})) to the following
discretized set of
momenta: $\{ \frac{1}{K}P^+, \frac{2}{K}P^+,  \frac{3}{K}P^+, \dots \}$.
Here, $P^+$ denotes the total light-cone momentum of a state,
and may be thought of as a fixed constant,
since it is easy to form a Fock basis that is already diagonal 
with respect
to the quantum operator $P^+$ \cite{pb85}. The integer $K$ is called
the `harmonic resolution', and $1/K$ measures the coarseness of our 
discretization --
we recover the continuum by taking the limit $K \rightarrow \infty$. 
Physically, $1/K$ represents the smallest 
positive\footnote{We exclude the zero mode $k^+=0$ in our analysis;
the massive spectrum is not expected to be affected by this 
omission \cite{bat98},
but there are issues concerning the light-cone vacuum that 
involve $k^+=0$ modes \cite{pin97a,mrp97}.}
unit of longitudinal momentum-fraction allowed for 
each parton in a Fock state.  

Of course, as soon as we implement the DLCQ procedure,
which is specified unambiguously by the harmonic resolution $K$,
the integrals appearing in the definitions for $Q^{\pm}_{\alpha}$
are replaced by finite sums, and the eigen-equation
$2 P^+ P^- |\Psi \rangle = M^2 |\Psi \rangle$ is reduced
to a finite matrix diagonalization problem. In this last
step, we use the fact that $P^-$ is proportional to the square
of any one of the two supercharges $Q^-_{\alpha}$, $\alpha=1,2$
(equation (\ref{superQminus})),
and so the problem of diagonalizing $P^-$ is 
equivalent to diagonalizing any one of the 
two supercharges $Q^-_{\alpha}$. As was pointed out
in \cite{sakai95}, this procedure yields a supersymmetric
spectrum for any resolution $K$.
In the present work, we are able to perform numerical diagonalizations
for $2 \leq K \leq 6$ with the help of Mathematica and a desktop PC.
 
In the next section,
we discuss the details of our numerical results.

\section{DLCQ Bound State Solutions}
\label{numericalresults}
We consider discretizing the light-cone supercharge
$Q^-_{\alpha}$ for a particular $\alpha \in \{1,2\}$, and
for the values $2 \leq K \leq 6$.
 For a given resolution $K$,
the light-cone momenta of partons 
in a given Fock state must be some positive integer 
multiple of $P^+/K$, where $P^+$ is the total light-cone momentum of
the state.

Of course,
the fact that we may choose any one of the two supercharges
to calculate the spectrum provides a strong test for
the correctness of our computer program, and consistency
of the DLCQ formulation. It turns out, however, that there are a few
surprises in store. First of all, the supersymmetry algebra
(eqns (\ref{superQplus})-(\ref{crossterm})) is certainly true
in a continuum space-time, but there is no obvious
reason to expect that
these relations should also hold exactly in a discretized version
of the theory.
From our numerical studies, however, we find that 
relations (\ref{superQplus}) and (\ref{crossterm}) are 
indeed exactly satisfied in the discretized theory.

A potential problem arises, however, in relation (\ref{superQminus}).
First of all, one finds that $Q_1^-$ and $Q_2^-$ do not
anti-commute: $\{Q_1^-,Q_2^-\} \neq 0$. However, 
this is not too surprising, since one can
trace this anomaly to the truncation of momenta (i.e. there is
a non-zero lower bound on $k^+$) following
from the DLCQ procedure. In particular, as we increase
the resolution, the non-zero matrix entries in  
$\{Q_1^-,Q_2^-\}$ become more and more sparsely distributed,
and we expect them to occupy a subset of measure zero in the   
continuum limit $K \rightarrow \infty$. This is substantiated
by direct inspection of the matrix $\{Q_1^-,Q_2^-\}$
for different values of $K$.

We also encounter a further anomaly when computing
the difference $(Q_1^-)^2 - (Q_2^-)^2$. 
According to relation (\ref{superQminus}),
this difference is precisely zero in the continuum, but 
in the discretized
theory, it is non-vanishing. As we discussed
above, this can be understood as an artifact of the
truncated momenta in the DLCQ formulation, and disappears
in the continuum limit $K \rightarrow \infty$. 

Nevertheless, we should worry at this stage about 
the definition of the light-cone Hamiltonian. 
Relation (\ref{superQminus}) suggests 
that we may define the DLCQ light-cone Hamiltonian
as the square of any one of the supercharges.
Because the difference $(Q_1^-)^2 - (Q_2^-)^2$ is non-vanishing
in the discretized theory,
we have two possible choices
for defining the light-cone Hamiltonian:
$P_1^- = \frac{1}{\sqrt{2}}(Q_1^-)^2$ or 
$P_2^- = \frac{1}{\sqrt{2}}(Q_2^-)^2$.
Surprisingly, after diagonalizing each of these
operators for different $K$, the spectrum of eigenvalues turns
out to be identical! This is certainly another 
attractive feature of DLCQ that deserves further study\footnote{
One suggestion is that these `discrepancies' in the operator algebra
are related to large gauge transformations arising from 
the light-like compactification in DLCQ, and are therefore expected to
vanish in the continuum limit $K \rightarrow \infty$. For finite
$K$, the operator `anomalies' we observed may be gauge equivalent 
to zero.}.

Of course, this implies that
the spectrum of the theory for finite $K$ is 
independent of the choice of supercharge, and therefore well defined.
It would be interesting to investigate whether other physical
observables are independent of the observed anomaly in
the supersymmetry algebra.

Let us begin with a discussion of massless states.
Firstly, for gauge symmetry U($N$),
the U(1) degrees of freedom explicitly decouple, 
and this  provides trivial examples
of massless states, which can be seen in the DLCQ 
analysis.  
Consequently, all the non-trivial
dynamics is contained in the SU($N$) gauge
theory. 
For $K=2$, the SU($N$) Fock space consists of
two parton states only. Moreover, since $Q^-_{\alpha}$ increases
or decreases the number of partons by one, it necessarily annihilates
all Fock states, and so all states are massless. 
However, for $K\geq 3$, determining
the existence (or not) of massless states in the SU($N$) 
theory turns out to be a highly non-trivial problem
involving the full dynamics of the Hamiltonian. 
We will therefore restrict our attention to the 
bound state spectrum of the SU($N$)
gauge theory.

The results of our DLCQ numerical diagonalization of $(Q^-_{\alpha})^2$
at resolution $K=3$ is summarized in Table \ref{K3masses}.  
To test our numerical
algorithm, we diagonalize (the square of) each supercharge, and find
the same spectrum -- which is consistent with supersymmetry.
\begin{table}[h!]
\begin{center}
\begin{tabular}{|c|c|}
\hline
\multicolumn{2}{|c|}{Bound State Masses $M^2$ for $K=3$ } \\
\hline
$M^2$ & Mass Degeneracy  \\
\hline
0 & $8+8$  \\
\hline
1.30826 & $4+4$ \\
\hline 
12.6273 & $4+4$  \\
\hline
22.0645 & $4+4$  \\
\hline 
\end{tabular}
\caption{SU($N$) bound state masses $M^2$ 
in units $g^2 N/\pi$ for resolution $K=3$ (six significant
figures). 
When expressed in these units, the masses are
independent of $N$ (i.e. there are no $1/N$
corrections at this resolution), and so these results are applicable
for any $N > 1$.
The notation `$4+4$'  above implies 
an exact 8--fold mass degeneracy in the DLCQ spectrum with
4 bosons and 4 fermions. In total, there are 20 bosons and 20 fermions. 
  \label{K3masses}}
\end{center}
\end{table}
Let us now consider resolution $K=4$. 
The results of our numerical diagonalizations are presented
in Table \ref{K4masses}. 
\begin{table}[h!]
\begin{center}
\begin{tabular}{|c|c|}
\hline
\multicolumn{2}{|c|}{Bound State Masses $M^2$ for $K=4$ } \\
\hline
$M^2$ & Mass Degeneracy  \\
\hline
0 & $32+32$  \\
\hline
1.20095 & $8+8$ \\
\hline 
4.00943 & $4+4$  \\
\hline
12.2424 & $4+4$  \\
\hline
12.2962 & $8+8$ \\ 
\hline 
15.0490 & $4+4$ \\
\hline
15.2822 & $4+4$ \\
\hline 
19.5028 & $8+8$ \\
\hline 
20 & $4+4$ \\
\hline
22.5321 & $4+4$ \\
\hline
23.1272 & $4+4$ \\
\hline 
28.6177 & $4+4$ \\
\hline
28.6955 & $4+4$ \\
\hline
\end{tabular}
\caption{SU($N$) bound state masses $M^2$ 
in units $g^2 N/\pi$ for resolution $K=4$ (six significant
figures).
When expressed in these units, the masses are
independent of $N$ (i.e. there are no $1/N$
corrections at this resolution), and so these results are applicable
for any $N > 1$. In total, there are 92 bosons and 92 fermions at this
resolution.
  \label{K4masses}}
\end{center}
\end{table}
If we express the masses in units $g^2 N/\pi$, then there are no
$1/N$ corrections at resolutions $K=3$ and $K=4$. However,
for $K \geq 5$, one sees $1/N$ effects in the spectrum.
In Table \ref{K5masses}, we list bound state masses for $N=3,10,100$
and $1000$ at resolution $K=5$. At this resolution, there are $472$
bosons and $472$ fermions.
\begin{table}[h!]
\begin{center}
\begin{tabular}{|c|c|c|c|}
\hline
\multicolumn{4}{|c|}{Bound State Masses $M^2$ for $K=5$, 
and $N=3,10,100,1000$ } 
\\
\hline
\multicolumn{4}{|c|}{$M^2$}  \\
\hline
$N=3$ & $N=10$ & $N=100$ & $N=1000$ \\
\hline
0.0 & 0.0 & 0.0 & 0.0  \\
\hline
0.0442062 & 0.0112824  & 0.00679546  & 0.00674981 \\
\hline 
 0.658859 & 0.634820 & 0.630485 &   0.630441 \\
\hline
1.13442 & 1.08578 & 1.08135  &   1.08131 \\
\hline
1.13442 &  1.11224 & 1.10995 &  1.10993 \\ 
\hline 
1.23157 &  1.56551 & 1.57314 &   1.57321 \\
\hline
1.29964 & 2.09691 & 2.17960 &  2.18043 \\
\hline 
1.55373  & 2.10814 & 2.17971 & 2.18043 \\
\hline 
1.76132 & 2.14535 & 2.18009 &  2.18043 \\
\hline
1.77999  & 2.14571 & 2.18009  & 2.18043 \\
\hline
\end{tabular}
\caption{SU($N$) bound state masses $M^2$ 
in units $g^2 N/\pi$ for resolution $K=5$ (six significant
figures), and for $N=3,10,100$ and $1000$. We have 
selected the lightest 10 states in each case, with mass degeneracy
$4+4$ for non-zero masses. Massless states have
degeneracy $92+92$. Note that if a state has degeneracy $8+8$, then we
include it twice (e.g. for $N=3$, there is a bound state
with mass squared $M^2=1.13442$ with degeneracy $8+8$). 
Convergence in the large $N$ limit is evident.
  \label{K5masses}}
\end{center}
\end{table}
Table \ref{K5masses} illustrates mass splittings that occur in
the spectrum as a result of $1/N$ interactions, which become 
increasingly important as we decrease $N$. For example, at $N=1000$,
there is an apparent degeneracy 
in the numerical spectrum at $M^2 = 2.18043$ 
which is visibly broken when $N=10$.

It turns out that these states
are easily identified  as weakly bound multi-particles
at large (but finite) $N$. To show this, note that 
bound states at $K=2$ are necessarily massless -- $M^2(K=2)=0$ --  while 
for $K=3$, 
the lightest non-zero mass state satisfies
$M^2(K=3) = 1.30826$. The mass squared $M^2(K=5)$ of a freely
interacting  {\em multi-particle}  
composed of one $K=2$ particle and one $K=3$ particle now
follows from simple kinematics \cite{grosshash}:
\begin{equation}
\frac{M^2(K=5)}{5} = \frac{M^2(K=3)}{3} + \frac{M^2(K=2)}{2},
\end{equation}
The result is $M^2(K=5) =  2.18043$, after inserting the observed values
for $M^2(K=3)$ and $M^2(K=2)$. Note that this value for $M^2(K=5)$
is a {\em prediction} for the mass of two freely interacting particles
at resolution $K=5$ (or equivalently, carrying $K=5$ units of
light-cone momentum). Thus, for large
enough $N$ (or for sufficiently small coupling $1/N$), 
we expect to see bound states approaching this 
two-free-body mass.
Table \ref{K5masses} confirms this prediction. 
Such predictions of multi-particle masses were also carried out
for the ${\cal N} = (1,1)$ model \cite{alp98}, and are 
a strong consistency test of the (typically complex) numerical 
algorithms adopted in this work.
  
Note, in general, that the $1/N$ interactions increase the 
masses of light particles, and decrease the mass of heavy particles 
(see Table \ref{K5masses}).
 
\begin{figure}[h]
\begin{center}
\epsfig{file=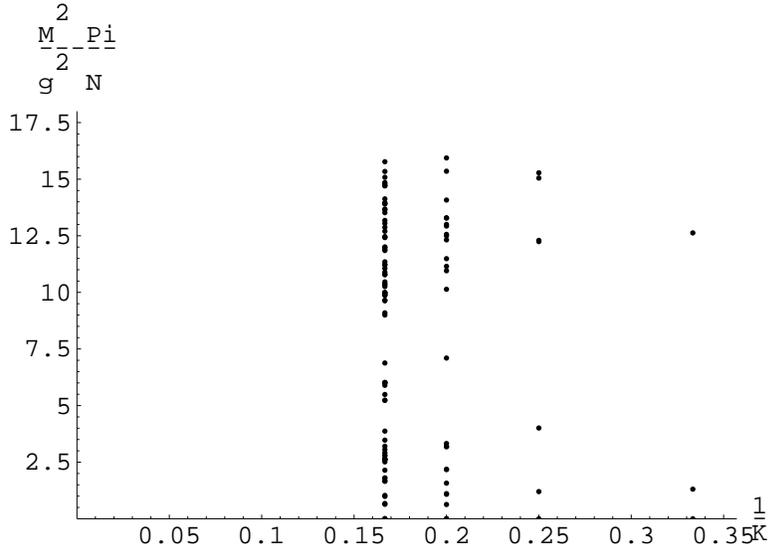}
\end{center}
\caption{{\small Bound State Masses $M^2$ (in units $g^2 N/\pi$)
versus $1/K$ for $N=1000$. We only plot masses satisfying $M^2 < 16$,
but there are many bound states above this for $K \geq 5$.} 
\label{one}}
\end{figure}

For $K=6$, there are over $4500$ states in the Fock basis. The 
resulting DLCQ spectrum for $N=1000$ appears in Figure \ref{one}, together
with bound state masses obtained at the lower resolutions
$K=3,4$ and $5$. It is apparent from this graph that as we increase
$K$ (i.e. as we move from right to left), the DLCQ spectrum seems 
to approach some dense subset of the positive real (vertical) axis.
One may infer that in the continuum limit $K \rightarrow \infty$,
the spectrum does indeed `fill up' the vertical axis, which is certainly
compatible with a recent study that suggested this theory
should be in a screening phase \cite{armoni}. 

As we pointed out earlier, decreasing $N$ introduces noticeable splittings 
in the spectrum\footnote{Some splittings are in fact present for
$N=1000$, but are not seen in the numerical spectrum since we quote
masses to only six significant figures.} which has the 
effect of smearing out the points in Figure \ref{one}. The 
qualitative features of the spectrum expected from a 
screening theory are therefore also present for smaller values 
of $N$.  

\medskip

So far, we have only discussed properties of the DLCQ spectrum --
such as bound state masses and their corresponding degeneracies -- but 
solving the DLCQ bound state equations also involves deriving
explicit numerical expressions for bound state {\em wave functions}.
This is of particular interest to us here since we would like to know
whether such a theory exhibits a mass gap or not.
In the context of our DLCQ analyses, this
involves establishing, in addition to the trivial
light-cone vacuum, the existence -- or not -- of a 
{\em normalizable} massless state in the continuum
theory $K \rightarrow \infty$. 
  
According to the literature, the ${\cal N}=(2,2)$
model is believed {\em not} to have 
a mass gap \cite{witt95}. We now outline how our numerical
results support such a claim\footnote{The ${\cal N} =(8,8)$
model, in contrast, was shown to have a mass gap \cite{alpII98}.}.

Firstly, at resolution $K=3$, and $N=1000$, one identifies a massless 
boson (and its superpartners) that has the form\footnote{
We choose not to normalize states to unity, since it is
very time consuming computationally when working at finite $N$, 
and not necessary 
when solving for spectra. A simpler
procedure is just to renormalize each Fock state
by $1/N^{(q/2)}$, where $q$ is the number 
of partons in the Fock state (implicitly implied
in this work). Then the {\em relative} size of each 
Fock state wave function 
indicates the relative importance of the Fock state to the
overall bound state.} 
\begin{equation}
   \mbox{tr}[a_1^{\dagger}(\frac{P^+}{3})a_2^{\dagger}
(\frac{2P^+}{3})]|0\rangle +
\mbox{tr}[a_2^{\dagger}(\frac{P^+}{3})a_1^{\dagger}
(\frac{2P^+}{3})]|0\rangle,
\label{K3soln}
\end{equation}
where $P^+$ is the total (fixed) momentum.
At resolution $K=4$, one identifies a massless boson
of the form
\begin{eqnarray}
 \lefteqn{ 0.497134 \times \mbox{tr}[a_1^{\dagger}(\frac{2P^+}{4})a_2^{\dagger}
(\frac{2P^+}{4})]|0\rangle \hspace{2mm} + \hspace{2mm}
  0.580827 \times \mbox{tr}[a_1^{\dagger}(\frac{P^+}{4})a_2^{\dagger}
(\frac{3P^+}{4})]|0\rangle} & & \nonumber \\ 
& &+ \hspace{3mm}
  0.495501\times \mbox{tr}[a_2^{\dagger}(\frac{P^+}{4})a_1^{\dagger}
(\frac{3P^+}{4})]|0\rangle \hspace{2mm}  + \hspace{2mm}
\mbox{additional Fock states},
\label{K4soln}
\end{eqnarray} 
where ``additional Fock states'' represents a superposition of 
two and four parton Fock states with amplitudes less than $0.25$
(typically, very small).
It is therefore natural to identify
the bound state solution above for $K=4$ 
with the $K=3$ solution [Eq.(\ref{K3soln})].
At $K=5$, something seems to go wrong; there are no  massless
states that may
be characterized as a superposition of predominantly two-parton states,
as in eqns (\ref{K3soln}) and (\ref{K4soln}). However, there
is a state with mass squared\footnote{
This is not a $1/N$ effect. If we let $N \rightarrow \infty$, 
the mass squared
$M^2$ does {\em not} converge to zero.} $M^2 = 0.0067$ ($N=1000$),
which has the explicit form
\begin{eqnarray}
 \lefteqn{ 0.52344 \times \mbox{tr}[a_1^{\dagger}(\frac{4P^+}{5})a_2^{\dagger}
(\frac{P^+}{5})]|0\rangle \hspace{2mm} + \hspace{2mm}
   0.468159 \times \mbox{tr}[a_1^{\dagger}(\frac{2P^+}{5})a_2^{\dagger}
(\frac{3P^+}{5})]|0\rangle} & & \nonumber \\ 
& &+ \hspace{3mm}
 0.468159 \times \mbox{tr}[a_1^{\dagger}(\frac{3P^+}{5})a_2^{\dagger}
(\frac{2P^+}{5})]|0\rangle \hspace{2mm}  + \hspace{2mm}
 0.52344 \times \mbox{tr}[a_2^{\dagger}(\frac{4P^+}{5})a_1^{\dagger}
(\frac{P^+}{5})]|0\rangle \nonumber \\
& & \hspace{2mm} + \hspace{2mm}
\mbox{additional Fock states},
\label{K5soln}
\end{eqnarray} 
where ``additional Fock states'' above represents a superposition of 
four parton Fock states with relatively small amplitudes. 
Evidently, it is natural to associate this bound state with 
the massless bound states (\ref{K3soln}) and (\ref{K4soln})
observed at lower resolutions. 

At this point, we would like to know whether this non-zero mass will
persist in the continuum
limit $K \rightarrow \infty$, or whether it is  
an artifact of the Fock state truncation enforced by the DLCQ
procedure. 

As it turns out, solving the DLCQ bound state equations at resolution 
$K=6$ reveals an exactly massless (bosonic)
solution that is essentially a superposition of two-parton
Fock states (as in  eqns (\ref{K3soln}),(\ref{K4soln}) and (\ref{K5soln})),
with wave functions that have approximately
the same shape, relative magnitude and sign as the
wave functions appearing at lower resolutions. 
The `glitch' in the spectrum observed at $K=5$, therefore, appears 
to be an artifact of the numerical truncation, although it would be 
desirable to probe larger values of $K$ (e.g. $K \geq 7$) to confirm
this viewpoint.

At any rate, we have identified a 
series of DLCQ solutions that is expected to converge in the limit 
$K \rightarrow \infty$ to a massless 
bound state. This continuum solution would rule out 
the possibility of a mass gap, in agreement with \cite{witt95}. 

We should remark at this point that the Fock state representation
of the lowest energy states in this model
appear to be significantly more complicated than
solutions found in other field theories with massless particles -- such as the 
t' Hooft pion, or Schwinger particle. A theory with complex
adjoint fermions studied relatively recently \cite{anp97,pin97} revealed
many massless states with constant wave function solutions.
In contrast, it turns out that any normalizable state in 
the (continuum) supersymmetric model studied here 
must be a superposition
of an infinite number of Fock states \cite{oleg}.
An analogous situation occurred in the model with
${\cal N}=(1,1)$ supersymmetry \cite{alp98}. 

Finally, we comment on the `string-like' nature of 
bound states that dominate the low energy spectrum. 
Although we focused on a massless state
composed of mainly two-parton Fock states,
one can always find a massless bound state dominated
by Fock states with an arbitrarily large number of partons
for sufficiently large $K$.
The structure of the low energy spectrum is similar, 
consisting of bound states of all lengths permitted by 
the truncation parameter $K$. Such qualitative features
of the spectrum were exhibited also in the 
${\cal N}=(1,1)$ supersymmetric model \cite{alp98,sakai95}.

\section{Discussion}
\label{conclusions}
To summarize, we have performed a detailed analysis of the DLCQ bound 
state spectrum of an ${\cal N}=(2,2)$ supersymmetric matrix model,
which may be heuristically derived by dimensionally
reducing ${\cal N}=1$ super-Yang-Mills from four to two space-time 
dimensions.
The gauge group is SU($N$), and we allow $N$ to be finite,
but arbitrarily large.

We discretize the light-cone supercharges via DLCQ, and
find that certain supersymmetry relations are exactly
satisfied even in the discretized formulation. In particular,
relations (\ref{superQplus}) and (\ref{crossterm}) hold
exactly in DLCQ. We find, however, that relation
(\ref{superQminus}) holds only approximately, although
the discrepancy diminishes as the resolution $K$ is increased.
As a consequence, the difference 
between the discrete light-cone supercharges $(Q_1^-)^2 - (Q_2^-)^2$
is non-zero. Surprisingly, however, the eigenvalues of $(Q_1^-)^2$
and $(Q_2^-)^2$ turn out to be {\em identical} at any resolution,
and so the DLCQ spectrum of the theory has an unambiguous definition
as the eigenvalues of either supercharge squared.
With this definition, the DLCQ spectrum turns out to be {\em exactly
supersymmetric
at any resolution} (see Tables \ref{K3masses} and \ref{K4masses},
for example). 
It would be desirable to understand why --- in the DLCQ
formulation --- the supersymmetry
operator algebra is only approximately satisfied, while
the spectrum itself appears to reflect unbroken supersymmetry.
Perhaps the observed anomaly in the algebra cancels for physical
observables\footnote{We might be able to redefine the supercharges
$Q_1^- \rightarrow R Q_1^- R^{-1}$, $Q_2^- \rightarrow S Q_2^- S^{-1}$,
for appropriate matrices $R,S$,
so that relation (\ref{superQminus}) holds exactly even in the
discretized theory. However, this would imply a non-vanishing result
for relation (\ref{crossterm}). Nevertheless, it would be tempting
to argue that this non-vanishing contribution 
(or `central charge') reflects the topology
induced by compactification of the light-like circle in DLCQ
\cite{yamawaki}.}, which would be consistent with the idea
that certain quantities are `gauge equivalent' to zero.

In Table \ref{K5masses}, we illustrate the dependence
of bound state masses on the number of gauge colors $N$.
Convergence is evident at large $N$ if we keep
$g^2 N$ constant. We also resolve 
mass splittings in the spectrum as a 
result of $1/N$ interactions (see Table \ref{K5masses}).
It appears that decreasing $N$ (i.e. increasing the strength
of the $1/N$ interactions) has the generic effect of decreasing
particle masses, except for very light particle states.
Note that there is no $N$ dependence of the spectrum
at resolutions $K=3$ and $4$ -- one needs to consider $K \geq 5$
to observe any variation with $N$.

In Figure \ref{one}, we plotted the DLCQ spectrum for resolutions
$K=3,4,5$ and $6$, and observed that the supersymmetric spectrum
approaches a dense subset of the positive real axis.
This is compatible with the recent claim that the theory is
in a screening phase \cite{armoni}. 

By carefully studying the Fock state content of certain
bound states at different resolutions, we argued for the existence
of a normalizable massless state in the continuum limit 
$K \rightarrow \infty$. A mass gap is therefore expected to be absent
in this theory.
  
We also observed that the low energy spectrum is dominated by
states with arbitrarily many partons --- a constituent picture 
involving few-parton Fock states is obviously an inadequate 
representation for capturing the full low energy dynamics of this model.

In the light of this highly complex bound state structure, it is 
tempting to suggest that we are probing a dynamical
system that might be more adequately (and simply) described by 
an effective field 
theory in higher dimensions\footnote{The presence of two transverse scalar
fields suggests a non-critical string theory in 
four dimensions.}.
Following the remarkable proposals of Matrix Theory 
and the AdS/CFT correspondence, it would be interesting to
pursue this speculation further. 

\medskip
\begin{large}
 {\bf Acknowledgments}
\end{large}

F.A. is grateful for hospitality bestowed
by the Max-Planck-Institute (Heidelberg)
where this work was begun.   
\appendix
\section{Appendix: Super-Yang-Mills in Four Dimensions}
\label{ymills4}
Let's start with ${\cal N}=1$ super Yang-Mills theory 
in 3+1 dimensions with gauge group U($N$) or SU($N$):
\begin{equation}
S_{3+1}=\int d^{4}x \hspace{1mm} \mbox{tr} \Bigg
(-\frac{1}{4} F_{\mu\nu}F^{ \mu\nu}+\frac{{\rm i}}{2} 
\bar{\Psi}\Gamma^{\mu}D_{\mu}\Psi\Bigg) , 
\label{EQ1}
\end{equation} 
where 
\begin{eqnarray}
F_{\mu\nu}&=&\partial_{\mu}A_{\nu}-\partial_{\nu}A_{\mu}
+{\rm i}g[A_{\mu},  A_{\nu}] , \\
D_{\mu}\Psi &=& \partial_{\mu}\Psi+{\rm i}g[A_\mu, \Psi],  
\end{eqnarray} 
and $\mu,\nu = 0,\dots,3$.
The Majorana spinor  $\Psi$ 
transforms in the adjoint representation of the gauge group. 
The (flat) space-time metric
$g_{\mu \nu}$ has signature $(+,-,-,-)$, and we adopt 
the normalization $\mbox{tr}(T^aT^b) = \delta^{a b}$ for
the generators of the gauge group.

The supersymmetry transformations 
\begin{eqnarray}
 \delta A_{\mu} & = & {\rm i} {\overline \epsilon} \Gamma_\mu \Psi \\
\delta \Psi & = & \frac{1}{2} F_{\mu \nu} \Gamma^{\mu \nu} \epsilon 
\end{eqnarray}
where $\Gamma^{\mu \nu} = \frac{1}{2} [\Gamma^\mu,\Gamma^\nu ]$
give rise to the following supercurrent:
\begin{equation}
J^{\mu} = \frac{{\rm i}}{2} \mbox{tr} \left(
\Gamma^{\rho \sigma} \Gamma^{\mu} F_{\rho \sigma} \Psi \right).
\label{supercurrent}
\end{equation}
In order to realize the four dimensional Dirac algebra  
$\{\Gamma_\mu, \Gamma_\nu\}=2g_{\mu\nu}$ in terms
of Majorana matrices, 
we use as building blocks the 
following three $2 \times 2$ real anti-commuting matrices: 
\begin{equation}
\epsilon_1=\left(\begin{array}{cc}
0 & 1\\
1 & 0 
\end{array}\right), \hspace{4mm}
\epsilon_2=\left(\begin{array}{cc}
0 & -1\\
1 & 0 
\end{array}\right), \hspace{4mm}
\epsilon_3=\left(\begin{array}{cc}
1 & 0\\
0 & -1 
\end{array}\right). \hspace{4mm}
\end{equation} 
We may now define four $4 \times 4$ pure-imaginary matrices
as tensor products of the above matrices:
\begin{equation}
\Gamma^0 = {\rm i} \epsilon_2 \otimes \epsilon_1 \hspace{6mm}
\Gamma^1 = {\rm i} \epsilon_1 \otimes {\boldmath 1} \hspace{6mm}
\Gamma^2 = {\rm i} \epsilon_3 \otimes {\boldmath 1} \hspace{6mm} 
\Gamma^3 = {\rm i} \epsilon_2 \otimes \epsilon_2 
\end{equation}
and it follows that these matrices satisfy 
$\{\Gamma_\mu, \Gamma_\nu\}=2g_{\mu\nu}$, as required. In our
numerical work, we use this particular representation.

To formulate the theory in light-cone coordinates,
it is convenient to introduce matrices 
\begin{equation}
\Gamma^{\pm} = 
\frac{1}{\sqrt{2}}(\Gamma^0 \pm \Gamma^3) \hspace{7mm}
\Gamma^I=\left(\begin{array}{cc}
{\rm i} \beta_I & 0 \\
0 & {\rm i} \beta_I 
\end{array}\right),
\end{equation}
where the two $2 \times 2$ real-symmetric matrices
$\beta_I$, $I=1,2$, are defined by writing 
$\beta_1 = \epsilon_1$,
and $\beta_2 = \epsilon_3$. 

It is now straightforward to verify that 
$P_L \equiv \frac{1}{2}\Gamma^+ \Gamma^-$ and 
$P_R \equiv \frac{1}{2}\Gamma^- \Gamma^+$ project out the
left and right-moving components of the four-component
spinor $\Psi$. Defining $\psi$ by a rescaling, $\Psi = 2^{1/4} \psi$,
we introduce left/right-moving spinors $\psi_{L,R}$  as follows:    
\begin{equation}
 \psi = \psi_R + \psi_L, \hspace{5mm}  \psi_R =  P_R \psi, \hspace{3mm}
\psi_L =  P_L \psi.
\end{equation}
This decomposition is particularly useful when working
with light-cone coordinates, since in the light-cone gauge
one can express the left-moving component $\psi_L$ in terms of
the right-moving component $\psi_R$ by virtue of the
fermion constraint equation. We will derive this result shortly.
In terms of the usual four dimensional Minkowski
space-time coordinates, the light-cone coordinates are given by
\begin{eqnarray}
 x^+ & = & \frac{1}{\sqrt{2}}(x^0 + x^3), \hspace{10mm}
\mbox{``time coordinate''} \\
 x^- & = & \frac{1}{\sqrt{2}}(x^0 - x^3), \hspace{10mm}
\mbox{``longitudinal space coordinate''}  \\ 
 {\bf x}^{\perp} & = & (x^1,x^2).
\hspace{13mm} \mbox{``transverse coordinates''} 
\end{eqnarray}
Note that the `raising' and `lowering' of the $\pm$ indices
is given by the rule $x^{\pm} = x_{\mp}$, 
while $x^I = -x_I$ for $I=1,2$,
as usual. It is now a routine task to demonstrate that
the Yang-Mills action (\ref{EQ1}) is equivalent to
\begin{eqnarray}
S_{3+1}^{LC} & = & \int dx^+ dx^- d{\bf x}^{\perp} \hspace{1mm}
 \mbox{tr} \left( \frac{1}{2}F_{+-}^2 + F_{+I}F_{-I} - \frac{1}{4}F_{IJ}^2
 \right. \nonumber \\
& & \hspace{20mm}
+ \hspace{1mm}
{\rm i} \psi_R^T D_+ \psi_R +   {\rm i}\psi_L^T D_- \psi_L 
     - {\rm i}\sqrt{2}\psi_L^T \epsilon_2 \beta_I D_I \psi_R \left. 
\frac{}{} \right),
\label{LCversion}
\end{eqnarray}
where the repeated indices $I,J$ are summed over $\{1,2\}$.
Some surprising simplifications follow if we now choose 
to work in the {\em light-cone gauge} $A^+ = A_- = 0$. In
this gauge $D_- \equiv \partial_-$, and so the (Euler-Lagrange)
equation of motion for the left-moving field $\psi_L$
is simply 
\begin{equation}
 \partial_- \psi_L = \frac{1}{\sqrt{2}} \epsilon_2 \beta_I D_I \psi_R, 
\label{fermioncon}
\end{equation}
which is evidently a non-dynamical constraint equation, since it
is independent of the light-cone time. We may therefore eliminate
any dependence on $\psi_L$ (representing unphysical 
degrees of freedom) in favor of $\psi_R$, which carries the
eight physical fermionic degrees of freedom in the theory.    
In addition, 
the equation of motion for the $A_+$ field yields
Gauss' law:
\begin{equation} 
\partial_{-}^2 A_{+}=\partial_{-}\partial_{I}A_{I}+gJ^+
\label{apluscon}
\end{equation} 
where $J^+={\rm i}[A_{I},\partial_{-}A_{I}]+2\psi_{R}^T\psi_{R}$, and
so the $A_+$ field may also be eliminated
to leave the two bosonic degrees of freedom $A_I$, $I=1,2$.
Note that the  two fermionic degrees of freedom
exactly match the 
bosonic degrees of freedom associated with the transverse
polarization of a four dimensional gauge field, which is of
course consistent
with the supersymmetry. We should emphasize that
unlike the usual 
covariant formulation of Yang-Mills, the light-cone formulation
here permits one to remove {\em explicitly} 
any unphysical degrees of freedom in the 
Lagrangian (or Hamiltonian); there 
are no ghosts, and supersymmetry is manifest.


%

\vfil

\end{document}